\documentclass[twocolumn,showpacs,preprintnumbers,amsmath,amssymb]{revtex4}

\usepackage{graphicx}
\usepackage{dcolumn}
\usepackage{bm}
\usepackage{braket}

\begin{document}

\preprint{APS/123-QED}

\title{Quantum phase with spontaneous translational symmetry breaking in an extended diamond chain}


\author{H. Yamaguchi$^{1,2}$, S. C. Furuya$^{3,4}$, Y. Tominaga$^{1}$, A. Matsuo$^{4}$, and K. Kindo$^{4}$}
\affiliation{
$^1$Department of Physics, Osaka Metropolitan University, Osaka 599-8531, Japan\\
$^2$Innovative Quantum Material Center (IQMC), Osaka Metropolitan University, Osaka 599-8531, Japan\\
$^3$Department of Liberal Arts, Saitama Medical University, Moroyama, Saitama 350-0495, Japan\\
$^4$Institute for Solid State Physics, the University of Tokyo, Chiba 277-8581, Japan\\
}

Second institution and/or address\\
This line break forced

\date{\today}

\begin{abstract}
We report the experimental realization of a spin-1/2 extended diamond chain in a verdazyl-Cu complex, where competing interactions and lattice distortions give rise to exotic quantum phases. 
The magnetic properties exhibit a zero-field energy gap and 1/2 magnetization plateau, which is explained by a dimer-monomer model.
Considering the effective interactions between the monomers, three types of dimer-dimer phases are expected to appear as the ground state, depending on the magnitude of the lattice distortions. 
By mapping to the nonlinear sigma model, three phases are distinguished topologically, and a symmetry-protected topological phase equivalent to the Haldane phase is identified.
Furthermore, a nontrivial magnetization is observed above the 1/2 plateau region, suggesting a gapped dimer phase accompanied by a spontaneous breaking of translational symmetry.
The discovery of this rare quantum state has broad implications for strongly correlated systems, topological matter, and quantum information science, where symmetry and topology play crucial roles.
\end{abstract}


\maketitle 
\section{INTRODUCTION}
Quantum spin systems serve as a rich platform for exploring exotic quantum phases, where the interplay of quantum fluctuations, symmetry, and topology gives rise to a variety of emergent phenomena. 
Among these, symmetry-protected topological (SPT) phases have attracted significant attention, as they exhibit nontrivial edge states protected by symmetries, with potential applications in quantum computation and topological quantum matter~\cite{spt1,spt2,haldane1,haldane2,haldane3}. 
The Haldane phase in spin-1 chains is a well-known example of an SPT phase, yet its realization in experimental systems remains a formidable challenge. 
Identifying SPT phases in diverse quantum spin models is therefore crucial for both fundamental physics and practical applications.
Simultaneously, spontaneous symmetry breaking (SSB) is a fundamental concept in physics, governing the emergence of ordered states from condensed matter to high-energy systems~\cite{nambu1960,anderson1984,tasaki2020}. 
In classical systems, translational symmetry breaking is widely observed in charge- and spin-density waves, but in quantum systems, strong fluctuations often suppress such ordering. 
A promising way to realize such quantum symmetry breaking lies in frustrated spin systems, where geometric frustration prevents conventional magnetic order and instead promotes quantum correlations. 
Frustration can induce effective interactions that stabilize symmetry-breaking phases, even in low-dimensional systems~\cite{kageyama1999,nishimoto2013}

The study of frustrated spin systems is a key area in condensed matter physics, providing a vast canvas for the exploration of emergent phenomena and novel quantum states~\cite{F1,F2,F3}. 
Within the intricate interplay between frustration and quantum effects in one-dimensional (1D) systems, unprecedented quantum phases emerge, shedding light on the fundamental principles governing quantum magnetism~\cite{1DF1,1DF2,1DF3}.
A prominent class of such systems involves spin networks composed of strongly interacting spin-dimer units.
In this context, the orthogonal-dimer chain has drawn considerable attention as a paradigmatic example in which geometric frustration and dimerization cooperate to produce magnetization plateaus and exotic quantum ground states.
The orthogonal-dimer chain is known to exhibit rich quantum behavior, including exact dimer ground states, an infinite sequence of magnetization plateaus, and even spontaneous translational symmetry breaking in certain parameter regimes~\cite{OD1,OD2,OD3,OD4,OD5}.
Although the coupling geometry in the present system differs from the orthogonal arrangement, it features competing dimer-like interactions and frustration that lead to a qualitatively similar physics.
These models collectively serve as fertile ground for understanding how frustration and dimerization can give rise to exotic symmetry-broken phases in low-dimensional quantum magnets.

Among various dimer-based systems, the diamond chain, characterized by its linear connection of triangular units, is notably recognized as another prototypical 1D frustrated model. 
The unique geometric arrangement of spins in the diamond chain gives rise to a distinctive interplay between quantum fluctuations and frustration. 
Previously, the spin-1/2 distorted diamond chain, which comprises inequivalent exchange interactions, has been extensively studied as a fundamental model.
Numerical studies predicted that a variety of exotic quantum states emerge depending on the delicate balance between three exchange interactions~\cite{takano, okamoto1,okamoto2,riron3,riron4}.
Experimentally, several copper-based compounds have been reported to form spin-1/2 distorted antiferromagnetic (AF) diamond chains~\cite{exp1,exp2,exp3,exp4,exp5,exp6}.
Among them, the azurite Cu$_3$(CO$_3$)$_2$(OH)$_2$ is a representative candidate material and has been intensively studied~\cite{exp1,exp2,theo1,theo2}.
In particular, high-field magnetization measurements of this material exhibit a 1/3 magnetization plateau, which has prompted further investigation into the quantum properties of the diamond chain~\cite{exp1}.

In this paper, we successfully synthesized single crystals of the verdazyl-based complex [$p$-Py-V-($p$-F)$_2$]$[$Cu(hfac)$_2]$ ($p$-Py-V-($p$-F)$_2$ = 3-(4-pyridinyl)-1,5-bis(4-fluorophenyl)-verdazyl, hfac = 1,1,1,5,5,5-hexafluoro-2,4-pentanedione). 
Molecular orbital (MO) calculations indicated the formation of a spin-1/2 chain with a diamond unit.
Magnetic properties exhibited a zero-field energy gap and 1/2 magnetization plateau, which can be explained by the full polarization of the Cu spin monomer with the singlet dimer composed of radical spins.
We quantitatively clarify that three types of dimer-dimer phases are expected to appear as a ground state depending on the magnitude of the lattice distortions. 
These phases are mapped to the nonlinear sigma model (NLSM)~\cite{NLSM1,NLSM2}, revealing a symmetry-protected topological phase equivalent to the Haldane phase~\cite{SPT}.
Furthermore, we find a nontrivial behavior in the magnetization curve above the 1/2 plateau phase.

\section{EXPERIMENTAL}
We synthesized [$p$-Py-V-($p$-F)$_2$]$[$Cu(hfac)$_2]$, the molecular structure of which is depicted in Fig. 1(a), by initially preparing $p$-Py-V-($p$-F)$_2$ using the conventional procedure~\cite{gosei}.
The subsequent synthesis of [$p$-Py-V-($p$-F)$_2$]$[$Cu(hfac)$_2]$ was accomplished following a previously reported procedure for verdazyl-based complexes~\cite{Mn_alt, tsukiyama}.
Recrystallization from a mixed solvent of CH$_2$Cl$_2$ and $n$-heptan yielded dark-brown crystals of [$p$-Py-V-($p$-F)$_2$]$[$Cu(hfac)$_2]$, as shown in Fig. 1(b). 

Single crystal X-ray diffraction was performed using a Rigaku XtaLAB Synergy-S instrument. 
To measure the magnetic susceptibility, we utilized a commercial SQUID magnetometer (MPMS, Quantum Design). 
The experimental results were corrected by considering the diamagnetic contributions calculated using Pascal's method.
Magnetization curves were measured using a non-destructive pulse magnet at the Institute for Solid State Physics (ISSP), University of Tokyo.
All experiments were performed using small, randomly oriented single crystals.

The MO calculations were performed using the UB3LYP method in the Gaussian 09 program package. 
The basis sets are 6-31G (between Cu and radical) and 6-31G($d$,$p$) (between radicals). 
The convergence criterion was 10$^{-8}$ hartree. 
To estimate the intermolecular magnetic interaction, we applied the evaluation scheme to multispin systems that have been studied previously using the Ising approximation~\cite{MOcal}.

\section{RESULTS}
\subsection{Crystal structure and spin model}
The crystallographic parameters of [$p$-Py-V-($p$-F)$_2$]$[$Cu(hfac)$_2]$ are provided in Table I.
In the molecular structure of [$p$-Py-V-($p$-F)$_2$]$[$Cu(hfac)$_2]$, the Cu atom is coordinated by a radical, resulting in a 5-coordinate environment, as shown in Fig.1(a).
Table II lists the bond lengths and angles of the Cu atom.
Both $p$-Py-V-($p$-F)$_2$ and Cu$^{2+}$ have spin-1/2.
The MO calculations revealed the presence of two AF and two ferromagnetic predominant exchange interactions.
These interactions are quantified as $J_{\rm{1}}/k_{\rm{B}}$ = $54$ K, $J_{\rm{2}}/k_{\rm{B}}$ = $-68$ K, $J_{\rm{3}}/k_{\rm{B}}$ = $41$ K, and $J_{\rm{4}}/k_{\rm{B}}$ = $-3$ K, which are defined within the Heisenberg spin Hamiltonian, given by $\mathcal {H} = J_{n}{\sum^{}_{<i,j>}}\textbf{{\textit S}}_{i}{\cdot}\textbf{{\textit S}}_{j}$.
$J_{\rm{1}}$ and $J_{\rm{4}}$ are intermolecular interactions between radical spins, as shown in Fig. 1(c).
The molecular pairs associated with these interactions are related by inversion symmetry.
$J_{\rm{2}}$ and $J_{\rm{3}}$ are interactions between the spins on the radicals and Cu atoms, as shown in Fig. 1(d).
Notably, the MO estimates tend to overestimate the interactions between verdazyl radicals and transition metals.
From our previous studies on the verdazyl-based complexes with similar coordination structures, the actual absolute values of $J_{\rm{2}}$ and $J_{\rm{3}}$ are expected to be approximately in the range of 10 K to 30 K~\cite{morota,frusthoneycomb,pPyVpCN}.
Furthermore, the intermolecular ferromagnetic interactions attributed to the molecular stacking of the verdazyl radicals, i.e., $J_{\rm{4}}$ in the present case, tend to be underestimated by a few K in MO calculations~\cite{3Cl4FV, 3Br4FV}.
Based on the MO estimates for verdazyl-based complexes, the relationship $J_{\rm{1}}$ ${\textgreater}$ ($|J_{\rm{2}}|$,$J_{\rm{3}}$) ${\textgreater}$ $|J_{\rm{4}}|$ is considered reliable.
These four interactions form a spin-1/2 1D chain with frustrated triangular couplings, as shown in Fig. 1(e).
Since this spin lattice has a diamond unit composed of $J_{\rm{2}}$, $J_{\rm{3}}$, and $J_{\rm{4}}$, similarities with the the orthogonal-dimer chain and the diamond chain can be expected.  
Considering the difference between the actual values and the MO evaluations for $J_{\rm{2}}$, $J_{\rm{3}}$, and $J_{\rm{4}}$, their actual values are expected to be close enough to enhance the frustration effect.

\begin{figure}[t]
\begin{center}
\includegraphics[width=20pc]{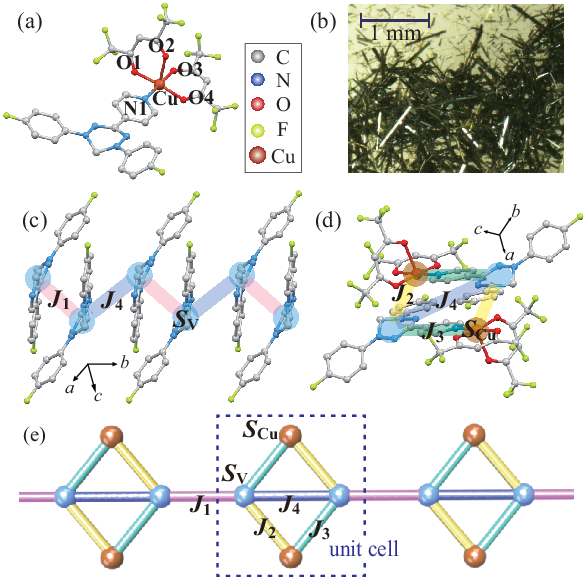}
\caption{(color online) (a) Molecular structure of [$p$-Py-V-($p$-F)$_2$]$[$Cu(hfac)$_2]$. Hydrogen atoms are omitted for clarity. (b) Single crystals of [$p$-Py-V-($p$-F)$_2$]$[$Cu(hfac)$_2]$. Crystal structure of [$p$-Py-V-($p$-F)$_2$]$[$Cu(hfac)$_2]$ forming (c) the chain along the $b$-axis and (d) the diamond unit. The blue and brown nodes represent the spin-1/2 of the radicals and Cu atoms, respectively. The thick lines represent the exchange interactions. (e) Spin-1/2 extended diamond chain composed of $J_{\rm{1}}$, $J_{\rm{2}}$, $J_{\rm{3}}$, and $J_{\rm{4}}$.
}
\end{center}
\end{figure}

\begin{table}
\caption{Crystallographic data of [$p$-Py-V-($p$-F)$_2$]$[$Cu(hfac)$_2]$.}
\label{t1}
\begin{center}
\begin{tabular}{lc}
\hline
\hline 
Formula & C$_{29}$H$_{16}$F$_{14}$CuN$_{5}$O$_{4}$\\
Crystal system & Monoclinic \\
Space group & $P2_{1}/n$ \\
Temperature (K) & 100(2) \\
$a$ $(\rm{\AA})$ & 15.0978(3) \\
$b$ $(\rm{\AA})$ & 9.4295(2)  \\
$c$ $(\rm{\AA})$ & 22.8709(4)  \\
$V$ ($\rm{\AA}^3$) & 3200.71(11) \\
$\beta$ (degrees) & 100.575(2) \\
$Z$ & 4 \\
$D_{\rm{calc}}$ (g cm$^{-3}$) & 1.718\\
Total reflections & 3749 \\
Reflection used & 3353 \\
Parameters refined & 478 \\
$R$ [$I>2\sigma(I)$] & 0.0704  \\
$R_w$ [$I>2\sigma(I)$] & 0.1836 \\
Goodness of fit & 1.064 \\
CCDC & 2325042\\
\hline
\hline
\end{tabular}
\end{center}
\end{table}

\begin{table}
\caption{Bond lengths ($\rm{\AA}$) and angles ($^{\circ}$) related to the Cu atom in [$p$-Py-V-($p$-F)$_2$]$[$Cu(hfac)$_2]$.}
\label{t1}
\begin{center}
\begin{tabular}{cc@{\hspace{1.5cm}}cc}
\hline
\hline
Cu--N1 & 2.02 & N1--Cu--O1 & 93.3 \\
Cu--O1 & 1.96 & O1--Cu--O3 & 86.9 \\
Cu--O2 & 2.23 & O3--Cu--O4 & 89.7 \\
Cu--O3 & 1.98 & O4--Cu--N1 & 89.9 \\
Cu--O4 & 1.93 & N1--Cu--O2 & 96.5 \\
 &  & O2--Cu--O3 & 84.7 \\
 &  & O3--Cu--N1 & 178.8 \\
 &  & O1--Cu--O2 & 88.6 \\
 &  & O2--Cu--O4 & 101.0 \\
 &  & O4--Cu--O1 & 169.6 \\
\hline
\hline
\end{tabular}
\end{center}
\end{table}

\subsection{Magnetization}
Figure 2(a) shows the temperature dependence of the magnetic susceptibility ($\chi$ = $M/H$) at 0.1 T.
A broad peak representing the AF correlations is found at approximately 5.0 K. 
In the low-temperature regime, below the broad peak, $\chi$ undergoes a significant decrease, indicating the existence of a nonmagnetic ground state separated from the excited states by an energy gap.
The temperature dependence of $\chi$$T$ shows a two-step decrease with decreasing temperature, as shown in the inset of Fig. 2(a). 
If a significant difference in the magnitude of exchange interactions exists, an energy separation occurs owing to the difference in the temperature region where the correlation becomes dominant, yielding a multistep change in $\chi$$T$~\cite{b26Cl2V,a26Cl2V,tominagaNi}. 
Accordingly, the observed two-step change in $\chi$$T$ is expected to reflect the differences in the exchange interactions, i.e., $J_{\rm{1}}$ ${\textgreater}$ ($|J_{\rm{2}}|$,$J_{\rm{3}}$).
Figure 2(b) shows the low-temperature behavior of $\chi$ at various magnetic fields. 
The gapped behavior with the rapid decrease is suppressed by the application of magnetic fields, indicating that the zero-field energy gap is evaluated to be approximately 1$-$3 T.   

Figure 2(c) shows the magnetization curves in pulsed magnetic fields.
Gapped behavior becomes prominent as the temperature decreases, which is consistent with the $\chi$ behavior observed in the low-temperature region.
The result at 1.4 K indicates a zero-field energy gap of $H_{\rm{c}}$$\approx$2.5 T, as shown in the inset of Fig. 2(c). 
We also found a distinct 1/2 plateau for fields ranging from approximately 13 and 45 T.
Considering the exchange interactions evaluated from the MO calculations, this plateau indicates the full polarization of $S_{\rm{Cu}}$ with the remaining $J_{\rm{1}}$ singlet dimer composed of radical spins, as described in the inset of Fig. 2(c).
The magnetic field at the end of the plateau is expected to correspond to the energy gap associated with the singlet dimer.
This indicates that the dominant exchange interaction $J_{\rm{1}}$, which stabilizes the singlet dimer, is on the order of 60 K, significantly larger than the other interactions.
Based on the isotropic $g$-value of 2.0 for organic radicals, the magnetization of 1.09 ${\mu}_{\rm{B}}/f.u.$ at the plateau suggests an average $g$-value of approximately 2.18 for $S_{\rm{Cu}}$.
Furthermore, we observe a nontrivial behavior around 52 and 55 T, indicating the emergence of a quantum phase with a spontaneous breaking of translational symmetry.


\begin{figure}
\begin{center}
\includegraphics[width=21pc]{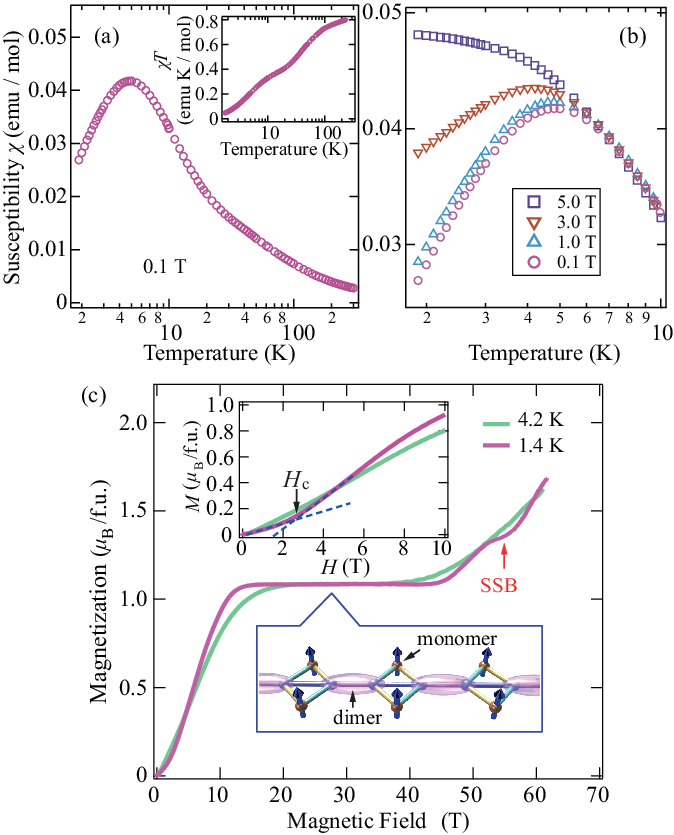}
\caption{(color online) (a) Temperature dependence of magnetic susceptibility ($\chi$ = $M/H$) of [$p$-Py-V-($p$-F)$_2$]$[$Cu(hfac)$_2]$ at 0.1 T.
The inset shows corrresponding $\chi T$ values.
(b) Temperature dependence of $\chi$ in the low-temperature region at various magnetic fields.
(c) Magnetization curve of [$p$-Py-V-($p$-F)$_2$]$[$Cu(hfac)$_2]$ at 1.4 K and 4.2 K.
The arrow shows the quantum phase accompanied by spontaneous symmetry breaking.
The illustrations describe the valence bond picture of the 1/2 plateau phase. 
The ovals and arrows represent the valence bond singlet of $S_{\rm{V}}$ dimer via $J_{\rm{1}}$ and $S_{\rm{Cu}}$ monomer polarized in the external field direction, respectively.
The upper inset shows magnetization curve in the low-field region, where the broken lines and arrow indicate the linear extrapolations and their intersection points, i.e, $H_{\rm{c}}$, respectively.
}\label{f2}
\end{center}
\end{figure}

\section{DISCUSSION}
\subsection{Candidate ground states}
We discuss the ground state of the anticipated spin-1/2 extended diamond chain with the coupling hierarchy $J_{\rm{1}} > (|J_{\rm{2}}|, J_{\rm{3}}) > |J_{\rm{4}}|$ at zero magnetic field.
The observed magnetization behavior suggests the formation of singlet dimers through the strongest AF interaction $J_{\rm{1}}$, which accounts for the wide 1/2 magnetization plateau.
As a consequence, the $S_{\rm{v}}$ spins forming the $J_{\rm{1}}$ dimers become effectively inactive at low temperatures ($T \ll J_{\rm{1}}/k_{\rm{B}}$), and the low-energy physics is governed by the remaining $S_{\rm{Cu}}$ spins.
Through second-order perturbation processes involving triplet excitations of the $J_{\rm{1}}$ dimers, effective interactions emerge between the $S_{\rm{Cu}}$ spins. 
These include $J_{r} = |J_{\rm{2}}|J_{\rm{3}}/J_{\rm{1}}$, $J_{l} = -|J_{\rm{2}}|J_{\rm{3}}/2J_{\rm{1}}$, $J_{d2} = J_{\rm{2}}^2/2J_{\rm{1}}$, and $J_{d3} = J_{\rm{3}}^2/2J_{\rm{1}}$, forming an effective spin-1/2 ladder with diagonal interactions, as illustrated in Fig. 3(a).
The effective spin Hamiltonian is written as  
\begin{multline}
\mathcal {H}_{\text{eff}} =  J_{r} {\sum_{j=1}^{N}} \textbf{{\textit S}}_{j,1} {\cdot} \textbf{{\textit S}}_{j,2}+J_{l} {\sum_{j=1}^{N}} {\sum_{\alpha=1,2}^{}} \textbf{{\textit S}}_{j,\alpha} {\cdot} \textbf{{\textit S}}_{j+1,\alpha}\\+J_{d2} {\sum_{j=1}^{N}} \textbf{{\textit S}}_{j,1} {\cdot} \textbf{{\textit S}}_{j+1,2}+J_{d3} {\sum_{j=1}^{N}} \textbf{{\textit S}}_{j,2} {\cdot} \textbf{{\textit S}}_{j+1,1},
\end{multline}
where $\textbf{{\textit S}}$ is the spin-1/2 operator.
This effective model captures the origin of the zero-field spin gap observed in the magnetization data, and provides a simplified microscopic picture valid in the strong-dimer limit ($J_{\rm{1}} \gg |J_{\rm{2}}|, J_{\rm{3}}, |J_{\rm{4}}|$), without aiming to reproduce the full field-dependent behavior of the original extended diamond chain.
Three types of ground states can be qualitatively distinguished depending on the difference in the absolute values of $J_{\rm{2}}$ and $J_{\rm{3}}$.
For $|J_{\rm{2}}|$=$J_{\rm{3}}$, the effective interactions have the relationship $J_r$ $\textgreater$ $|J_{l}|$=$J_{d1}$=$J_{d2}$, resulting in the formation of singlet dimers via $J_r$, as illustrated in Fig. 3(b).
Similarly, when $|J_{\rm{2}}|$ $\gg $ $J_{\rm{3}}$, singlet dimers are formed by the dominant AF interaction $J_{d2}$ (Fig.3(c)); whereas, when $J_{\rm{3}}$ $\gg $ $|J_{\rm{2}}|$, singlet dimers are formed by the dominant AF interaction $J_{d3}$ (Fig.3(d)).
Considering that the actual absolute values of $J_{\rm{2}}$ and $J_{\rm{3}}$ are expected to be approximately in the range of 10 K to 30 K, the effective interactions are expected to be on the order of a few K, which is consistent with the experimental results showing the peak of $\chi$ at $\sim$5 K and the energy gap of $\sim$2.5 T.

\begin{figure*}
\begin{center}
\includegraphics[width=40pc]{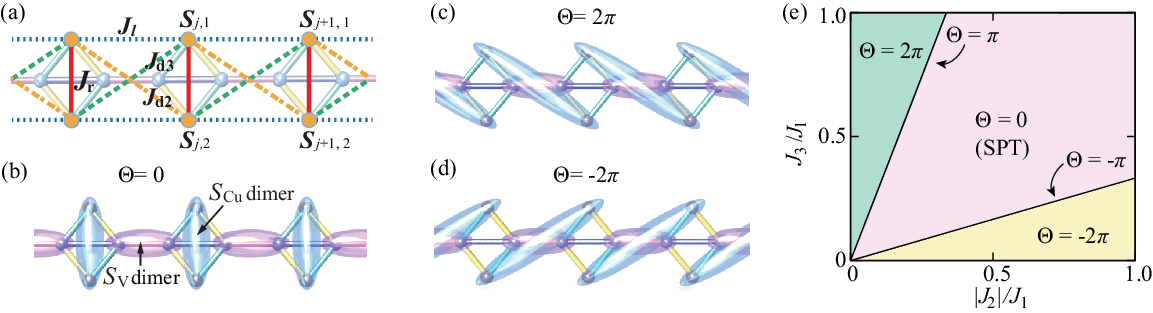}
\caption{(color online) (a) Effective spin ladder model with diagonal couplilngs on the extended diamond chain at $T$$\ll$$J_{\rm{1}}/k_{\rm{B}}$.
Valence bond pictures of the ground state for (b) $|J_{\rm{2}}|$=$J_{\rm{3}}$, (c)  $|J_{\rm{2}}|$ $\gg $ $J_{\rm{3}}$, and (d) $J_{\rm{3}}$ $\gg $ $|J_{\rm{2}}|$. 
The purple and blue ovals represent the valence bond singlets of $S_{\rm{V}}$ via $J_{\rm{V1}}$ and $S_{\rm{Cu}}$ via each dominant effective interaction, respectively.
(e) Schematic phase diagram in the spin-1/2 extended diamond chain on $|J_{\rm{2}}|$/$J_{\rm{1}}$-$J_{\rm{3}}$/$J_{\rm{1}}$ plane.
The solid lines denote the phase boundaries with $\Theta={\pm}\pi$.
$\Theta=0$ phase is an SPT and topologically distinguished from $\Theta={\pm}2\pi$ phases. 
}\label{f4}
\end{center}
\end{figure*}

\subsection{Topology of the quantum phases}
We examine the ground state phase diagram through a mapping to the NLSM, which allows the identification of the topology of the quantum phases. 
By introducing two fields, $\textbf{{\textit n}}(x)$ and $\textbf{{\textit l}}(x)$, we describe the spin vector as 
\begin{equation}
\textbf{{\textit S}}_{j,\alpha} =(-1)^{\alpha}S\textbf{{\textit n}}(x)+a_{0}\textbf{{\textit l}}(x),
\end{equation}
where $S=1/2$, $a_{0}$ is the spin-lattice constant along the leg direction, and $\textbf{{\textit n}}(x)$ and $\textbf{{\textit l}}(x)$ satisfy the following constraints~\cite{NLSM2}:
\begin{equation}
S^{2}\textbf{{\textit n}}(x)^{2}+a_{0}^{2}\textbf{{\textit l}}(x)^{2}=S^2, \quad \textbf{{\textit n}}(x){\cdot}\textbf{{\textit l}}(x)=0.
\end{equation}
Using the spin vector, the effective spin Hamiltonian is written as
\begin{multline}
\mathcal {H}_{\text{eff}} \approx {\int}dx\left[\frac{gv}{2}\left(\textbf{{\textit l}}-\frac{\Theta}{4\pi}\partial_{x}\textbf{{\textit n}}\right)+\frac{v}{2g}(\partial_{x}\textbf{{\textit n}})^2\right],
\end{multline}
where the coupling $g$, the velocity $v$, and the topological angle $\Theta$ are given by 
\begin{equation}
\begin{aligned}
g=\frac{1}{S}\frac{|J_{2}|+J_3}{\sqrt{|J_{2}|J_3}},\quad
v=2Sa_{0}\frac{|J_{2}|+J_3}{J_1}\sqrt{|J_{2}|J_3},\\
\Theta=4{\pi}S\frac{J_3-|J_{2}|}{J_3+|J_{2}|}.
\end{aligned}
\end{equation}
By introducing an operator canonically conjugate to $\textbf{{\textit n}}$, the kinetic term of Lagrangian $\mathcal {L}_{\text{kin}}$ can be derived~\cite{NLSM2}.
Then, considering $\mathcal {H}_{\text{eff}}$ as the potential term, the complete Lagrangian $\mathcal {L}$ is given by $\mathcal {L}_{\text{kin}}-\mathcal {H}_{\text{eff}}$.
In imaginary time ($t\rightarrow-i\tau $), $\mathcal {L}$ corresponds to $-\mathcal {\Tilde{H}}_{\text{eff}}$, leading to
\begin{multline}
\mathcal {\Tilde{H}}_{\text{eff}}=\frac{v}{2g}{\int}dx\left(\frac{1}{v^2}(\partial_{\tau}\textbf{{\textit n}})^2+(\partial_{x}\textbf{{\textit n}})^2\right)\\
-i\Theta{\int}\frac{dx}{4\pi}\textbf{{\textit n}}{\cdot}\partial_{\tau}\textbf{{\textit n}}\times\partial_{x}\textbf{{\textit n}},
\end{multline}
where the second term determines the topology in the partition function.
$\Theta=2\pi$ and $-2\pi$ correspond to $J_{2}=0$ and $J_{3}=0$, respectively, and $\Theta=0$ corresponds to  $|J_{2}|=J_{3}$.
These topological values identify the three dimer-dimer phases in Figs. 3(b)-3(d).
Furthermore, $\Theta=\pi$ and $-\pi$, corresponding to $J_{3}=3|J_{2}|$ and $J_{3}=|J_{2}|/3$, respectively, give two phase boundaries, as shown in Fig. 3(e).

We now explain why the $\Theta=0$ phase corresponds to an SPT phase, whereas the $\Theta=\pm 2\pi$ phases are topologically trivial in Fig. 3(e), despite the opposite convention~\cite{SPT}.
Whether the ground state of the spin-1/2 system is topological crucially depends on the choice of the unit cell.
When an odd number of singlet bonds extend across unit cells, the ground state realizes an SPT phase, whereas an even number of such bonds leads to a topologically trivial state.
This situation is analogous to the Su–Schrieffer–Heeger model of topological insulators.
Under the present unit cell (Fig.1(e)), the $\Theta=0$ phase corresponds to an SPT state, as a single singlet bond extends across adjacent unit cells. 
In contrast, the $\Theta={\pm}2\pi$ phases are topologically trivial, since two singlet bonds are formed between neighboring unit cells.
This delicate choice of the unit cell is decisive in determining whether the ground states are topological or trivial. 
Importantly, however, the essential distinction between the $\Theta=0$ phase and the $\Theta={\pm}2\pi$ phases remains independent of this choice. 
The former is unambiguously distinguished from the latter two by a protecting symmetry, namely the $D_2 \simeq Z_2 \times Z_2$ symmetry, in direct analogy with the $S=1$ Haldane phase~\cite{spt1,spt2}.


\subsection{Nontrivial behavior above the 1/2 plateau region}
Finally, we discuss the possible origin of the nontrivial behavior observed in the magnetization curve above the 1/2 plateau region.
In the higher field region beyond the plateau phase, $S_{\rm{Cu}}$ is fully polarized and does not have sufficient degrees of freedom to modify the ground state.
Therefore, we consider only the singlet state $\ket{S} = (\ket{\uparrow \downarrow}-\ket{\downarrow \uparrow})/\sqrt{2}$ and one of the triplet states $\ket{T_{\rm{1}}} = \ket{\uparrow \uparrow}$ formed by the $S_{\rm{V}}$ dimer near the end of the plateau phase.
The first-order perturbation treatment of $J_{\rm{2}}$, $J_{\rm{3}}$, and $J_{\rm{4}}$ gives the effective interactions between the the effective spin $\textbf{{\textit s}}$ with eigenstates $\ket{\uparrow} = \ket{T_{\rm{1}}}$ and $\ket{\downarrow} = \ket{S}$.
Thus, the effective spin Hamiltonian $\mathcal {H}_{\rm{eff2}}$ is written as
\begin{multline}
\mathcal {H}_{\text{eff2}} \approx J_{\rm{NN}}{\sum_{j=1}^{N}}(s_{j}^{x}s_{j+1}^{x}+s_{j}^{y}s_{j+1}^{y}-\frac{1}{2}s_{j}^{z}s_{j+1}^{z})\\-h_{\rm{eff}}{\sum_{i}}s_{i}^{z}+\rm{const.},
\end{multline}
where $J_{\rm{NN}}$=$|J_4|/2$ and $h_{\rm{eff}}$ = $g{\mu}_{\rm{B}}H$-$J_1$+$|J_4|/2$.
This spin Hamiltonian describes the spin-1/2 XXZ chain with the effective field $h_{\rm{eff}}$.
The 1/2 plateau and saturated phases correspond to $\ket{\downarrow}$ and $\ket{\uparrow}$, respectively, and a phase transition to the Tomonaga-Luttinger liquid occurs at $h_{\rm{eff}}$ = 0.
Furthermore, the consideration of the higher-order perturbations can lead to a next-nearest-neighbor interaction in the spin-1/2 XXZ chain~\cite{NNN}. 
If the next-nearest-neighbor interaction is strong enough, a dimer phase with a spin gap is expected to occur at $h_{\rm{eff}}$ = 0 accompanied by a spontaneous breaking of translational symmetry~\cite{dimer}.
The next-nearest-neighbor interaction introduces the geometrical frustration to the effective spin chain and results in the dimerization of spins on the nearest-neighbor bond in the ground state to lower the local energy on the triangle formed by $\bm{s}_{j-1}$, $\bm{s}_{j}$, and $\bm{s}_{j+1}$.
This dimer state is smoothly connected to the completely dimerized state of the Majumdar-Ghosh model~\cite{r1,r2}.
The Majundar-Ghosh model gives a special point to the phase diagram, where the above-mentioned local energy is proportional to $(\bm{s}_{j-1}+\bm{s}_j+\bm{s}_{j+1})^2$.
The ground state is thus spontaneously dimerized to minimize $(\bm{s}_{j-1}+\bm{s}_j+\bm{s}_{j+1})^2$.
Importantly, from the perspective of the Oshikawa–Yamanaka–Affleck criterion~\cite{oya}, the emergence of a magnetization plateau above 1/2 is forbidden unless the ground state breaks translational symmetry and enlarges the magnetic unit cell. 
In our model, such a plateau is not allowed under the original symmetry, and therefore, the observation of the nontrivial magnetization above the 1/2 plateau phase strongly suggests the realization of a spontaneously dimerized phase. 
This provides further support for interpreting the observed phase as a symmetry-broken gapped state induced by frustrated next-nearest-neighbor interactions.

\section{Summary}
To summarize, we successfully synthesized single crystals of [$p$-Py-V-($p$-F)$_2$]$[$Cu(hfac)$_2]$, a verdazyl-based complex.
MO calculations revealed the presence of four types of exchange couplings, resulting in a spin-1/2 extended diamond chain composed of Cu and radical spins.
We found a zero-field energy gap of about 2.5 T, a distinct 1/2 magnetization plateau for fields of about 13 and 45 T, a nontrivial quantum phase around 52 and 55 T.
We examined the ground state of the expected spin-1/2 extended diamond chain by the mapping on to the effective spin ladder model with diagonal couplings and reveled that three types of ground states can be qualitatively distinguished depending on the difference in the absolute values of the exchange interactions. 
By utilizing the NLSM, three dimer-dimer phases were topologically distinguished, and an SPT phase equivalent to the Haldane phase was identified.
We proposed that the quantum gapped state induced by the frustration is a strong candidate for the origin of the nontrivial high-field phase accompanied by SSB.
The present results open a new avenue of research focusing on diamond topology in the field of condensed matter physics.

\begin{acknowledgments}
We thank S. Shimono for valuable discussions.
A part of this work was performed under the interuniversity cooperative research program of the joint-research program of ISSP, the University of Tokyo.
\end{acknowledgments}

The data that support the findings of this article are openly available~\cite{ripo}.



\end{document}